\documentclass[a4paper,11pt]{article}  
  
\usepackage{amsfonts,amssymb}  

\def\G{\Gamma}    

\newcommand\plb[3]{{\it Phys.\ Lett.\ }{\bf B #1} (#2) #3}

\newcommand\prd[3]{{\it Phys.\ Rev.\ }{\bf D #1} (#2) #3}

\newcommand\jhep[3]{{\it J. High Energy Phys.\ }{\bf #1} (#2) #3}

\newcommand\prep[3]{{\it Phys.\ Rept.\ }{\bf #1} (#2) #3}

\newcommand\cmp[3]{{\it Commun.\ Math.\ Phys.\ }{\bf #1} (#2) #3}

\newcommand\npb[3]{{\it Nucl.\ Phys.\ }{\bf B #1} (#2) #3}

\newcommand\ap[3]{{\it Ann.\ Phys.\ (NY) }{\bf #1} (#2) #3}

\newcommand\prl[3]{{\it Phys.\ Rev.\ Lett.\ }{\bf #1} (#2) #3}

\newcommand\nc[3]{{\it Nuovo Cim.\ }{\bf #1} (#2) #3}

\begin{document}
\rightline{MPI-Pht/2002-66}

\vskip 0.8 truecm
\Large
\bf 

\centerline{Comments on the Equivalence between}
\centerline{Chern-Simons Theory and Topological}
\centerline{Massive Yang-Mills Theory in 3D}

\normalsize 
\rm

\vskip 0.5 truecm

\Large
\centerline{Andrea Quadri\footnote{E-mail: {\tt quadri@mppmu.mpg.de}}}

\normalsize
\vskip 0.3 truecm

\centerline{Max-Planck-Institut f\"ur Physik (Werner-Heisenberg-Institut)}
\centerline{F\"ohringer Ring 6, D80805 M\"unchen, Germany}
\vskip 0.2 truecm

\centerline{{\em and}}

\vskip 0.2 truecm
\centerline{INFN, Sezione di Milano}
\centerline{via Celoria 16, I20133 Milano, Italy}

\vskip 0.3 truecm
\bf
\centerline{Abstract}

\rm 

\begin{quotation}
The classical formal 
equivalence upon a redefinition of the gauge connection
between Chern-Simons theory and topological massive Yang-Mills theory
in three-dimensional Euclidean space-time
is analyzed at the quantum level within the BRST formulation of the Equivalence
Theorem. The parameter controlling the change in the gauge connection 
is the inverse $\lambda$ of the topological mass.
The BRST differential associated with the gauge connection
redefinition is derived and the corresponding Slavnov-Taylor (ST) 
identities are proven to be anomaly-free.
Hence they can be restored order by order in the loop
expansion by a recursive choice of non-invariant counterterms.
The Green functions of local operators constructed only from the 
($\lambda$-dependent) transformed
gauge connection, as well as those of BRST invariant operators,
are shown to be independent of the parameter $\lambda$, 
as a consequence of the validity of the ST identities.
The relevance of the antighost-ghost
fields, needed to take into account at the quantum level
the Jacobian of the change in the gauge connection,
 is analyzed. Their r\^ole in the identification of
the physical states of the model within conventional perturbative
gauge theory is discussed.
It is shown that they prove to be essential in keeping
the correspondence between the degrees of freedom of the theory
at $\lambda = 0$ (Chern-Simons theory) and at $\lambda \neq 0$.
\end{quotation}

\leftline{Key-words: BRST quantization, BRST symmetry, Chern-Simons theory}

\section{Introduction}

It has been observed some time ago \cite{sorella1,sorella2} that 
it is possible to recast the action of the 
classical three-dimensional topological  massive Yang-Mills theory
in flat Euclidean space-time
in the form of a pure Chern-Simons action via a non-linear 
redefinition of the gauge connection.
One can prove \cite{sorella1, sorella2} that the transformed
gauge field occurring into the Chern-Simons action
can be expressed as a formal power series
in the inverse $\lambda=\frac{1}{m}$ of the topological mass $m$:
\begin{eqnarray}
\hat A_\mu = A_\mu + \sum_{n=1}^\infty \lambda^n \theta^n_\mu(A) \, ,
\label{i1}
\end{eqnarray}
where $A_\mu$ is the gauge connection of topological massive Yang-Mills
theory and $\theta^n_\mu(A)$ is
the coefficient of $\hat A_\mu$ of order $n$ in the $\lambda$-expansion.
$\hat A_\mu$ is non-local since the R.H.S. of eq.(\ref{i1})
does not reduce to a polynomial. However, it turns out that
each coefficient $\theta^n_\mu(A)$ 
is local and covariant, being constructed only from the
field strength of $A_\mu$  and its covariant derivatives.
In particular, this implies that the transformed gauge field 
$\hat A_\mu$ is  again a gauge connection.

It is worthwhile noticing that this classical formal equivalence 
does not hold only for the three-dimensional 
topological massive Yang-Mills theory. It also applies
\cite{sorella2,Barnich:vg} to any local gauge-invariant
three-dimensional 
Yang-Mills-type action, defined as an arbitrary integrated local
gauge-invariant polynomial with zero ghost number,
built only with the gauge field strength and its covariant
derivatives. That is, at the classical level one can obtain
any local gauge-invariant three-dimensional Yang-Mills-type action (including those
which are not power-counting renormalizable) by starting
from the three-dimensional Chern-Simons action and then performing
a gauge field redefinition within the space of gauge connections.
This in turn provides a strong geometrical characterization
\cite{sorella2} of
three-dimension classical Yang-Mills-type action functionals: they 
can all be obtained
by evaluating the Chern-Simons functional at a suitable point
in the space of gauge connections which are formal power series
in the parameter $\lambda$.
In \cite{Barnich:vg} a complete cohomological analysis of this
property was given, establishing within the Batalin-Vilkovisky
formalism the rigidity of (non-Abelian) Chern-Simons theories.

The problem of whether this classical equivalence can also be extended
at the quantum level is yet an open issue. 
In this paper we will address this question within 
the BRST formulation of the Equivalence Theorem (ET) \cite{ET}.
We will consider for the sake of definiteness the equivalence
between three-dimensional Chern-Simons theory and topological 
massive Yang-Mills theory.
Our technique can however be applied as well to the more general
class of Yang-Mills-type actions.

In order to implement the classical equivalence at the quantum level
one needs to take into account the contribution from the Jacobian
of the field  redefinition in eq.(\ref{i1}) \cite{ET}.
This can be done by introducing
a suitable set of antighost-ghost fields $(\bar c_\mu,c_\mu)$, carrying the
same Lorentz-vector index as $A_\mu$.

We show that 
the introduction of these ghost-antighost fields
allows to gauge-fix the gauge symmetry of the three-dimensional
Chern-Simons classical action by using a Lorentz-covariant gauge
without introducing the standard Faddeev-Popov 
ghost and antighost scalar fields.

The resulting classical action turns out to be invariant under a new BRST
differential $\tilde s$. 
$\tilde s$ is
an extension of the gauge BRST differential $s$, under which the ungauged
classical action of three-dimensional Chern-Simons theory is invariant,
 combined with
the on-shell Equivalence Theorem BRST differential  \cite{ET} $\delta$,
taking into account the effects of the field redefinition in eq.(\ref{i1}).

The absence of the conventional scalar Faddeev-Popov (FP) 
ghost-anti\-ghost fields
constitutes an important difference with respect to the standard 
BRST gauge-fixing procedure. 
In fact 
it turns out that the local and covariant implementation 
of the Jacobian of the gauge connection redefinition in eq.(\ref{i1}) 
by means of the fields $(\bar c_\mu,c_\mu)$ cannot be performed
if the conventional BRST gauge-fixing procedure  for a Lorentz-covariant
gauge has already been carried out. This is due to the fact that 
in this case a degeneracy arises 
in the two-point functions in the ghost-antighost
sector of the classical action, which prevents the definition of the
propagators in the sector with non-zero ghost number.

The BRST differential 
$\tilde s$ is nilpotent only modulo the equations of motion
for $\bar c_\mu$. Hence we must resort to the full Batalin-Vilkovisky
\cite{gomis} formalism to obtain the Slavnov-Taylor (ST) identities
associated with $\tilde s$. In particular, the classical action
fulfilling the ST identities will acquire a non-linear dependence
on the antifield $\bar c^{\mu *}$.

We will prove 
by purely algebraic arguments 
that these ST identities can be restored
to all orders in the loop expansion.
Let us remark here that the transformed action is no more of the
power-counting renormalizable type.
Hence we cannot apply the 
Quantum Action Principle (QAP) \cite{qap1,qap2,qap3,qap4,Piguet:er}, 
which only holds for power-counting
renormalizable theories, in order to guarantee that 
the breaking of the ST identites is equal to the insertion
of the sum of a finite set of local operators of bounded dimension.

The latter property is of the utmost importance in 
the Algebraic Renormalization of gauge theories 
\cite{Piguet:er}. Indeed, if the breaking of the
ST identities can be characterized as the insertion of the sum of
a finite set
of local operators with bounded dimension, its first non-vanishing
order $\Delta^{(n)}$ in the loop expansion must reduce to a local polynomial
in the fields, the antifields and their derivatives. 
Moreover, from purely algebraic properties one can derive
a set of consistency conditions for $\Delta^{(n)}$, known as 
the Wess-Zumino consistency conditions \cite{Piguet:er,Wess:yu},
which identify $\Delta^{(n)}$ as an element of the kernel
of a suitable nilpotent differential operator, given by
the classical linearized ST operator \cite{Piguet:er}.
One can then characterize $\Delta^{(n)}$ by means of the powerful
techniques of cohomological algebra 
in the space
of local functionals \cite{Piguet:er,Barnich:2000zw}, 
to which $\Delta^{(n)}$ belongs.

The QAP cannot be extended on general grounds to non power-counting
renormalizable theories \cite{ET,Stora2000}.
We will then assume that a regularization scheme has been adopted 
fulfilling the so-called Quasi-Classical Action Principle (QCAP) \cite{ET,Stora2000}.
The fulfillment of the QCAP only implies that the first non-vanishing
order in the loop expansion of the breaking of the ST identities
is a local formal power series in the fields, the antifields and
their derivatives, without any reference to the all orders behaviour
of the ST breaking term.

In sharp contrast with the QAP, no characterization of the full breaking
of the ST identities to all orders in the loop expansion is possible
by means of the QCAP. Despite this fact, in the present case the QCAP
can be combined with the Wess-Zumino consistency condition 
\cite{Piguet:er,Wess:yu}
for the
ST breaking terms 
in order
to characterize 
on purely algebraic 
grounds
all possible ST breaking terms $\Delta^{(n)}$ which appear
at the first non-vanishing order $n$ in the loop expansion.
One can then recursively prove that the ST identities can be restored 
order by  order in the loop expansion by a suitable choice of non-invariant 
counterterms, due to the fact that the cohomology of the 
linearized classical ST operator
${\cal S}_0$ is empty in the space of local formal power series with 
ghost number $+1$, to which $\Delta^{(n)}$ belongs.

We will then show that the Green functions of strictly $\tilde s$-invariant
local operators are $\lambda$-independent, as a consequence of the fulfillment
of the ST identities.
The ST identities also imply that  
 the Green functions of the gauge field $\hat A_\mu(A,\lambda)$
and of those local composite operators only depending on $\hat A_\mu$
are  $\lambda$-indepen\-dent.
We remark that these operators are $\tilde s$-invariant
only modulo the equation of motion
for $\bar c_\mu$.

Such a selected set of Green functions 
can be equivalently computed from the quantum effective action
at $\lambda=0$ and from the quantum effective action at $\lambda \neq 0$.
This translates at the quantum level the classical equivalence
between three-dimensional Chern-Simons theory and topological massive
Yang-Mills theory in flat Euclidean space-time.

We stress that this property only holds if the antighost-ghost fields
$(\bar c_\mu,c_\mu)$ are taken into account in the quantization
of the model. They actually play an essential r\^ole in the identification
of the asymptotic physical degrees of freedom.

We refer here to a pure perturbative
analysis of the physical spectrum. As is known \cite{Guadagnini:1989kr}, 
the topological
nature of three-dimensional Chern-Simons theory prevents 
the existence
of any physical state in the sense of the conventional perturbative framework of gauge theories, with the exception of the vacuum.
On the contrary, ordinary topological massive Yang-Mills theory, without
the inclusion of the antighost-ghost fields $(\bar c_\mu, c_\mu)$,
does possess local excitations \cite{top_mass}.

In our approach 
the correspondence between the quantum theory at $\lambda=0$ and
at $\lambda \neq 0$ at the level of the asymptotic states is
preserved by virtue of the structure of $\tilde s$ and 
 the introduction of the antighost-ghost fields $(\bar c_\mu,c_\mu)$,
trivializing the cohomology of the asymptotic BRST operator $Q$
in the sector with zero ghost number.
This could in turn shed some light on the relevance at the 
asymptotic level
of the 
fields $(\bar c^\mu,c^\mu)$, which were 
originally introduced \cite{ET}
 as a tool to implement  in the path-integral
in a local and covariant way
the effect 
of the Jacobian of the field redefinition in eq.(\ref{i1}).

\vskip 0.5 truecm

The paper is organized as follows.
In sect.~\ref{sec2} we discuss the classical equivalence
between three-dimensional Chern-Simons theory and three-dimensional
topological massive Yang-Mills theory.
In sect.~\ref{sec3} we present the BRST formulation of the
Equivalence Theorem as applied to the field redefinition in eq.(\ref{i1}).
We discuss the gauge-fixing condition and we introduce the
antighost and ghost fields $(\bar c^\mu,c^\mu)$ associated
with the Jacobian of the field redefinition in eq.(\ref{i1}).
We also define the BRST differentials $s$, $\delta$ and $\tilde s$ and
derive the classical ST identities of the model.
In sect.~\ref{sec4} we show that the ST identities can be recursively restored
to all orders in the loop expansion by a suitable choice of non-invariant
counterterms. In sect.~\ref{sec5} we analyze the consequences of the
ST identities for the Green functions of the theory and establish
the $\lambda$-independence of the connected Green functions of any BRST 
invariant local operators, as well as of  those local
operators only depending on $\hat A(A,\lambda)$. 
We notice that the latter are BRST-invariant only modulo the equation of motion
for $\bar c^\mu$. Locality is always understood in the sense of local formal power
series.
In sect.~\ref{sec6} we analyze the physical states
of the original and transformed theory within the framework
of conventional perturbative field theory and prove that
no physical states are present, with the exception of the vacuum.
Hence the correspondence between the perturbative physical spectrum of
the theory at $\lambda=0$ (Chern-Simons theory) and at
$\lambda \neq 0$ is preserved.
Finally conclusions are presented in sect.~\ref{sec7}.

\section{The classical equivalence}\label{sec2}

We follow the conventions of \cite{sorella2}. The classical action of topological massive
Yang-Mills theory is given by
\begin{eqnarray}
S_{YM}(A) + S_{CS}(A)
\label{e1}
\end{eqnarray}
where $S_{YM}(A)$ is the Yang-Mills action 
\begin{eqnarray}
S_{YM}(A) = \frac{1}{4m} {\rm Tr} \int d^3x \, F_{\mu\nu}F^{\mu\nu} 
\label{e2}
\end{eqnarray}
and $S_{CS}(A)$ denotes the Chern-Simons term
\begin{eqnarray}
S_{CS} (A) = \frac{1}{2} {\rm Tr} \int d^3x \, \epsilon^{\mu\nu\rho} (A_\mu \partial_\nu A_\rho
+ \frac{2}{3} g A_\mu A_\nu A_\rho ) \, .
\label{e3}
\end{eqnarray}
$m$ in eq.(\ref{e2}) is the topological mass \cite{top_mass}, while
$g$ is the coupling constant of the model.
$A^\mu$ takes values in the Lie algebra $\mathfrak{G}$ of a compact
Lie group $G$:
\begin{eqnarray}
A_\mu = A_\mu^a T^a \, ,
\label{e3_1}
\end{eqnarray}
where the generators $T^a$ are assumed to be anti-hermitian.
The field strength $F_{\mu\nu}$ is given by
\begin{eqnarray}
F_{\mu\nu} = \partial_\mu A_\nu - \partial_\nu A_\mu + g [A_\mu,A_\nu] \, .
\label{e3_1_1}
\end{eqnarray}
$A_\mu$ has mass dimension $1$, while the parameters $g$ and $m$
have mass dimension $0$ and $1$ respectively.
We work in flat Euclidean space-time.

It has been shown in \cite{sorella1, sorella2} that 
the classical action in eq.(\ref{e1}) can be obtained from the
pure Chern-Simons action in eq.(\ref{e3}) by a suitable field
redefinition:
\begin{eqnarray}
S_{CS}(\hat A) = S_{YM}(A) + S_{CS}(A) \, ,
\label{e6}
\end{eqnarray}
where
\begin{eqnarray}
\hat A_\mu = A_\mu + \sum_{n=1}^\infty \frac{1}{m^n} \theta^n_\mu(D,F) \, .
\label{e4}
\end{eqnarray}
In the above equation the coefficients $\theta^n_\mu(D,F)$ are local and covariant,
being constructed only with the field strength $F_{\mu\nu}$ and
the covariant derivative
\begin{eqnarray}
D_\mu (\cdot) = \partial_\mu(\cdot) + g[A_\mu,\cdot] \, .
\label{e5}
\end{eqnarray}
We find it convenient to denote by $\lambda$ the inverse of $m$:
\begin{eqnarray}
\lambda = \frac{1}{m} \, .
\label{e7}
\end{eqnarray}
The field redefinition in eq.(\ref{e4}) is a formal power series
in $\lambda$. 

$S_{CS}(\hat A)$ is invariant under
\begin{eqnarray}
s \hat A_\mu = (D_{\hat A})_\mu \omega \, , 
\label{e8}
\end{eqnarray}
In the above equation the notation $(D_{\hat A})_\mu$ means
that the covariant derivative is computed with respect to the
gauge field $\hat A_\mu$. 
$\omega$ is the Lie-algebra valued classical ghost associated with
the BRST differential $s$.

The requirement that $s$ is nilpotent yields
\begin{eqnarray}
s \omega = -\frac{g}{2} \{ \omega,\omega \} \, .
\label{e9_1}
\end{eqnarray}
By setting
\begin{eqnarray}
s A_\mu = D_\mu \omega 
\label{e9}
\end{eqnarray}
we obtain
\begin{eqnarray}
s \hat A_\mu (A,\lambda) = (D_{\hat A(A,\lambda)})_\mu \omega \, ,
\label{e10}
\end{eqnarray}
due to the fact that the coefficients $\theta^n_\mu$ are covariant.
In eq.(\ref{e10}) $\hat A(A,\lambda)$ is understood as the formal power series
given in eq.(\ref{e4}).
Eq.(\ref{e10}) means that $\hat A_\mu(A,\lambda)$ is a connection.
As a consequence, $S_{CS}(\hat A(A,\lambda))$ is invariant under 
the BRST transformation in eq.(\ref{e9}).
This is what one should expect, since
the R.H.S. of eq.(\ref{e6}) is also
invariant under the transformation in eq.(\ref{e9}).

It has been observed \cite{sorella2,Barnich:vg} that any Yang-Mills-type
action (defined as an arbitrary integrated local invariant 
polynomial with vanishing ghost number, built only with the field
strength $F_{\mu\nu}$ and its covariant derivatives) can actually be obtained
by evaluating the Chern-Simon functional in eq.(\ref{e6})  on a suitable
gauge connection $\hat A_\mu$ of the form in eq.(\ref{e4}).
This in turn provides a strong geometrical characterization
\cite{sorella2} of
three-dimension classical Yang-Mills-type action functionals: they 
can all be obtained
by evaluating the Chern-Simons functional at a suitable point
in the space of gauge connections which are formal power series
in the parameter $\lambda$.
In \cite{Barnich:vg} a complete cohomological analysis of this
property was given, establishing within the Batalin-Vilkovisky
formalism the rigidity of (non-Abelian) Chern-Simons theories.

For the sake of definiteness we will analyze in the following
sections the equivalence between the three-dimensional Chern-Simons action
and massive topological Yang-Mills theory. We remark however that our technique
can also be applied in a straightforward way to the more general
situation where the equivalence between the Chern-Simons action and
an arbitrary Yang-Mills-type action is considered.

\section{BRST formulation of the gauge connection redefinition}
\label{sec3}

The equivalence discussed in sect.\ref{sec2} holds at the classical level.
The technique discussed in \cite{ET} provides a prescription, based on the
Slavnov-Taylor (ST) identities, to implement the field redefinition in eq.(\ref{e4})
at the quantum level.

In particular one needs to take into account the Jacobian of the
transformation in eq.(\ref{e4}) by introducing a suitable set of 
ghost and antighost fields \cite{ET}. 
We apply here the on-shell formalism of the Equivalence Theorem \cite{ET},
thus avoiding the introduction of auxiliary Lagrange multiplier fields,
which turn out to be unnecessary in the present case.
We stress that from the point of view of the
perturbative expansion the parameter $\lambda$ is treated here as an external
classical constant source.
We first define
\begin{eqnarray}
\Sigma & = & S_{CS}(\hat A(A,\lambda)) 
             \nonumber \\
       & = & S_{YM}(A) + S_{CS}(A) \, .
\label{ne2}
\end{eqnarray}
In order to quantize the model a gauge-fixing term must be introduced.
For this purpose we consider the following action:
\begin{eqnarray}
\Sigma' = \Sigma -\frac{\alpha}{2} {\rm Tr} \int d^3x \, 
\left ( \partial \hat A(A,\lambda) \right )^2 \, .
\label{ne2_bis}
\end{eqnarray}
$\Sigma'$ is no more $s$-invariant, due to the Lorentz-covariant gauge-fixing
in the R.H.S. of eq.(\ref{ne2_bis}).
We will comment on the choice of the gauge-fixing condition in eq.(\ref{ne2_bis})
later on in this section.

By following the ET on-shell quantization
prescription given in \cite{ET}
we include the contribution from the Jacobian of the transformation in eq.(\ref{e4})
by introducing a set of antighost-ghost fields $(\bar c_\mu, c_\mu)$.
$\bar c_\mu$ is $s$-invariant, while 
$c_\mu$ transforms as follows under $s$:
\begin{eqnarray}
s c_\mu = -g \{ c_\mu, \omega \} \, .
\label{m1}
\end{eqnarray}
The BRST differential $\delta$ associated with the field redefinition in eq.(\ref{e4}) is given by
\begin{eqnarray}
&& \delta \bar c_\mu = -\left . \frac{\delta \Sigma'}{\delta \hat A^\mu} \right |_{\hat A^\mu=\hat A^\mu(A,\lambda)} \,, 
~~~~~~ \delta A_\mu = c_\mu \, , ~~~~~~ \delta c_\mu = 0 \, ,  \nonumber \\  
&& \delta \lambda = \chi \, ,  ~~~~~ \delta \chi = 0 \, .
\label{new1_bis}
\end{eqnarray}
The full BRST differential $\tilde s$ is
\begin{eqnarray}
\tilde s = s + \delta \, .
\label{new2_bis}
\end{eqnarray}
For convenience we gather here the $\tilde s$-transformations
of the fields and the external sources 
$\lambda,\chi,\omega$:
\begin{eqnarray}
&& \tilde s A_\mu = D_\mu \omega + c_\mu \, , ~~~~
\tilde s \omega = -\frac{g}{2} \{ \omega,\omega \} \, , ~~~~
\tilde s \lambda = \chi \, , ~~~~ \tilde s \chi =0 \, ,
\nonumber \\
&& \tilde s c_\mu = -g \{ c_\mu, \omega \} \, , ~~~~ 
\tilde s \bar c_\mu = - \left . \frac{\delta \Sigma'}{\delta \hat A_\mu}
\right |_{\hat A_\mu = \hat A_\mu(A,\lambda)} \, .
\label{new2_ter}
\end{eqnarray}
Then the classical action
\begin{eqnarray}
\Sigma_0 & = & \Sigma' + {\rm Tr} \int d^3x \, \bar c^\mu \tilde s \hat A^\mu \nonumber \\
& = &  \Sigma' + {\rm Tr} \int d^3x \, \bar c^\mu \left ( \partial_\mu \omega + g[\hat A_\mu, \omega] \right) \nonumber \\
&   & + {\rm Tr} \int d^3x \, \bar c^\mu \left ( \frac{\delta \hat A^\mu}{\delta A ^\rho} c^\rho 
                                                          +\frac{\delta \hat A^\mu}{\delta \lambda} \chi \right ) 
\label{new2}
\end{eqnarray}
is $\tilde s$-invariant. 
The term in the last line of eq.(\ref{new2}) takes into account the Jacobian of the field redefinition.
We remark that 
$\omega$ is an external anti-commuting classical source.
It does not need to be promoted to a ghost field, since
$\Sigma'$ in eq.(\ref{ne2_bis}) is already gauge-fixed.
This constitutes an important difference with the standard Faddeev-Popov
BRST gauge-fixing procedure for gauge theories \cite{Weinberg:kr}.
In fact the addition of a conventional Lorentz-covariant
gauge-fixing term
\begin{eqnarray}
{\rm Tr} \int d^3x \, s \left ( \bar c ( \frac{\alpha}{2} B - \partial A ) \right ) =
{\rm Tr} \int d^3x \, \left ( \frac{\alpha}{2}B^2 - B \partial A 
- \partial^\mu \bar c D_\mu c \right )
\label{convFP}
\end{eqnarray}
by means of the set of scalar Faddeev-Popov fields $\bar c,c$
and the Nakanishi-Lautrup multiplier field $B$,
transforming as follows under $s$
\begin{eqnarray}
&& s A_\mu = D_\mu c \, , ~~~~ sc = -\frac{g}{2} \{ c,c \} \, , 
\nonumber \\
&& s \bar c = B \,  , ~~~~ s B = 0 \, , 
\label{convBRST}
\end{eqnarray}
would lead
to a degeneracy in the two-point functions in the 
 ghost-antighost sector of the classical action, hence
preventing the construction of the ghost propagators.

The r\^ole of the antighost-ghost fields $(\bar c_\mu,c_\mu)$
in the definition of the asymptotic physical states
of the model within the conventional framework
of perturbative field theory will be analyzed in sect.~\ref{sec6}.

\par

$\lambda$ and $\chi$ are constant external classical sources,
with $\chi$ anti-commuting.

We can assign the ghost number as follows: $A_\mu$ and $\lambda$
have ghost number zero, $\bar c^\mu$ has ghost number $-1$, 
$c^\mu$ has ghost number $+1$. $\omega$ and $\chi$ have ghost number $+1$.
With these assignments $\Sigma_0$ in eq.(\ref{new2}) has ghost number zero.

The mass dimension of $\bar c^\mu$ is $2$, the mass dimension of 
$c^\mu$ is $1$. $\lambda$, $\chi$ and $\omega$ have mass dimension zero.

We remark that 
$\tilde s$ is nilpotent on $\bar c^\mu$ only modulo the equation of motion for 
${\bar c}_\mu$:
\begin{eqnarray}
\tilde s^2 \bar c^\mu = - 
\frac{\delta^2 \Sigma'}{\delta \hat A_\mu \delta \hat A_\nu}
\tilde s \hat A_\nu = - 
\frac{\delta^2 \Sigma'}{\delta \hat A_\mu \delta \hat A_\nu}
\frac{\delta \Sigma_0}{\delta \bar c^\nu} \, .
\label{new2.1}
\end{eqnarray}
 Therefore the full Batalin-Vilkovisky formalism is needed
\cite{gomis} in order to write the ST identities associated
with $\tilde s$. The classical 
action will acquire a non-linear dependence on the
antifield $\bar c^*_\mu$. 

We consider the following classical action, constructed according
to the prescription given in \cite{ET} for the on-shell version
of the Equivalence Theorem:
\begin{eqnarray}
\G^{(0)}  & = & \Sigma_0
                  \nonumber \\
              &   & + {\rm Tr} \, \int d^3x \, \sum_{j=1}^\infty (-1)^j 
                         \frac{1}{j!} (\bar c^{\mu_1})^* \dots (\bar c^{\mu_j})^* 
                         \frac{\delta^{(j)} \Sigma'}{\delta \hat A_{\mu_1} \dots \delta \hat A_{\mu_j}} \, .
	          \nonumber \\
\label{e11}
\end{eqnarray}
$\G^{(0)}$ is a local formal power series in the antifield
$({\bar c}^\mu)^*$. This follows as a consequence of the fact that 
$\Sigma_0$ in eq.(\ref{new2}) is a local formal power series in the field
$A_\mu$, due to the gauge-fixing term entering into eq.(\ref{ne2_bis}).

$\G^{(0)}$ satisfies the following ST identities:
\begin{eqnarray}
\!\!\!\!\! {\cal S}(\G^{(0)}) & = & {\rm Tr} \int d^3x \, 
\left ( 
(\partial_\mu \omega + g[A_\mu,\omega] + c_\mu)
\frac{\delta \G^{(0)}}{\delta A_\mu}
       +\frac{\delta \G^{(0)}}{\delta (\bar c^\mu)^*}\frac{\delta \G^{(0)}}{\delta {\bar c}_\mu} \right . \nonumber \\
&& \left . 
 ~~~~     
-g \{c_\mu,\omega \} \frac{\delta \G^{(0)}}{\delta c_\mu}
-\frac{g}{2} \{ \omega,\omega \} \frac{\delta \G^{(0)}}{\delta \omega}
\right ) +\chi \frac{\partial \G^{(0)}}{\partial \lambda} = 0 \, .
\label{st1}
\end{eqnarray}
The corresponding linearized ST operator ${\cal S}_0$ is given by
\begin{eqnarray}
{\cal S}_0 & = & {\rm Tr}
\int d^3x \, 
\left ( 
(\partial_\mu \omega + g[A_\mu,\omega] + c_\mu)
\frac{\delta }{\delta A_\mu} 
 +\frac{\delta \G^{(0)}}{\delta (\bar c^\mu)^*}\frac{\delta}{\delta\bar c^\mu}  
+ \frac{\delta \G^{(0)}}{\delta {\bar c}_\mu} \frac{\delta}{\delta(\bar c^{\mu*})}  \right . \nonumber \\
& & ~~~~~~~~~~~~~ \left . 
-g \{c_\mu,\omega \} \frac{\delta}{\delta c_\mu}
-\frac{g}{2} \{ \omega,\omega \} \frac{\delta}{\delta \omega}\right )
    +\chi \frac{\partial}{\partial \lambda} \, .
\label{linsti}
\end{eqnarray}
We notice that $\lambda$ forms together with its partner $\chi$ a 
${\cal S}_0$-doublet. This in turn will allow to control the dependence of the Green functions of the model on $\lambda$.
The propagators for $A_\mu$ and $(\bar c^\mu,c^\mu)$, computed
from $\G^{(0)}$, both exist,
as it can be easily checked.
The classical action $\G^{(0)}$ together 
with the ST identities in eq.(\ref{st1}) is the starting point for the
quantization of the model.
\section{Quantization}\label{sec4}

In this section we show that 
the ST identities in eq.(\ref{st1}) can be restored
to all orders in the loop expansion. 
A formal proof of the fact that the ST identities in eq.(\ref{st1}) 
are anomaly-free at the quantum level is needed because
the inclusion of the gauge BRST differential $s$ into $\tilde s$
prevents a direct application of the results given in \cite{ET}
on the recursive fulfillment of the Equivalence Theorem ST identities,
associated with the pure Equivalence Theorem 
BRST differential $\delta$.

We first remark that
the Quantum Action Principle \cite{qap1,qap2,qap3,qap4,Piguet:er} does not apply,
since we are dealing with a non-power counting renormalizable theory.
Let us comment on this point further.
For  power-counting renormalizable theories the QAP characterizes
 the possible breaking of the ST identities, to all orders in perturbation theory,
 as the insertion of a finite sum of local operator with bounded dimension.
Consequently, if such an insertion were zero up to order $n-1$,
 at the $n$-th order it must reduce to a local polynomial in the fields
 and the external sources and their derivatives with bounded dimension.
This follows from the locality of the insertion and the topological
 interpretation of the $\hbar$-expansion as a loop-wise expansion.

One might notice that all what is needed to carry out the recursive analysis
 of the ST breaking terms by using cohomological and 
 algebraic methods is that part
 of the QAP, saying that at the first non-vanishing order
 in the loop expansion the ST breaking term is a local polynomial
 in the fields and the external sources and their derivatives.

It has then been proposed~\cite{Stora2000} that this consequence
of the QAP could be extended to non power-counting renormalizable
theories in the form of the so-called 
Quasi-Classical Action Principle, stating that
\begin{quotation}
\noindent
{\em
In the loop-wise perturbative expansion the first non-vanishing order
 of ST identities is a classical local integrated formal power
 series in the fields and external sources and their derivatives.}
\end{quotation}
We remark that the QCAP does not allow to characterize the possible
breaking of ST identities to all orders in the loop expansion,
independently of the non-invariant counterterms and the normalization
conditions chosen, as the QAP does.

Although its
 plausibility, no satisfactory proof is available on general grounds
 for the QCAP.
Thus from now on we will assume that it is fulfilled by the regularization
 used to construct the vertex functional $\G$.
Moreover, we point out that we do not require that
 the breaking term is polynomial:
 loosing the power-counting entails that no bounds on the dimension
 can in general be expected.

Since we assume the QCAP, we can now use the power of cohomological
 algebra to constrain the possible ET ST breaking terms and show
 that the ET ST identities can always be restored by a suitable order by order
 choice of non-invariant counterterms.

We suppose that the ST identities have been fulfilled up to order $n-1$
in the loop expansion:
\begin{eqnarray}
{\cal S}(\G)^{(j)} = 0 \, , ~~~~ j=0,1,\dots,n-1 \, .
\label{a1}
\end{eqnarray}
By virtue of the Quasi-Classical Action Principle 
the $n$-th order ST breaking term $\Delta^{(n)}$, given by
\begin{eqnarray}
\Delta^{(n)} \equiv {\cal S}(\G)^{(n)} \, , 
\label{a2}
\end{eqnarray}
is a local formal power series in the fields, the external sources
and their derivatives with ghost number $+1$. Moreover it satisfies 
the Wess-Zumino consistency condition \cite{Piguet:er,Wess:yu}
\begin{eqnarray}
{\cal S}_0 (\Delta^{(n)}) = 0 \, , 
\label{a3}
\end{eqnarray}
where ${\cal S}_0$ is the linearized classical ST operator given 
in eq.(\ref{linsti}).

We are going to show by purely algebraic methods that 
eq.(\ref{a3}) implies
\begin{eqnarray}
\Delta^{(n)} = {\cal S}_0 (-\Xi^{(n)}) 
\label{a4}
\end{eqnarray}
for a local formal power series $\Xi^{(n)}$ with ghost number zero.

Then the $n$-th order ST identities can  be restored by adding 
the non-symmetric counterterm functional
 $-\Xi^{(n)}$ to the $n$-th order vertex functional.
In order to prove eq.(\ref{a4}) we will first derive
some general results on the cohomology $H({\cal S}_0,{\cal F})$ 
of the linearized ST operator ${\cal S}_0$ in the space
of Lorentz-invariant integrated local formal power series
spanned by $A_\mu,\bar c^\mu, {\bar c}^{\mu*}, c^\mu,\omega$ and $\lambda,\chi$ and their derivatives.
At the first stages of this analysis we do not impose any restrictions
on the ghost number  of the functionals belonging to ${\cal F}$. In the end
we will specialize to the sector with ghost number $+1$, to which
$\Delta^{(n)}$ in eq.(\ref{a2}) belongs. 

Let us remark first that  
$(\lambda,\chi)$ form a set of doublets under ${\cal S}_0$,
since
\begin{eqnarray}
{\cal S}_0(\lambda) = \chi \, , ~~~~ {\cal S}_0(\chi) = 0 \, .
\label{a5_1}
\end{eqnarray}
$(\lambda,\chi)$ is  a coupled doublet,
in the sense that the counting operator
\begin{eqnarray}
{\cal N} = \lambda \frac{\partial}{\partial \lambda} +
           \chi \frac{\partial}{\partial \chi}
\label{a6}
\end{eqnarray}
does not commute with ${\cal S}_0$. 
Doublets can be handled by the powerful methods of homological
perturbation theory \cite{Henneaux:ig,Barnich:db,Barnich:mt,Barnich:2000zw}.
It can be proven \cite{Barnich:db,Barnich:mt,doppietti} that the cohomology of 
any nilpotent differential $\rho$ 
in the space of local formal integrated power series ${\cal P}$ 
spanned by a set of variables $\{\varphi,z,w\}$
and their derivatives
is isomorphic to the
cohomology of the restriction 
${\rho '}$
of 
$\rho$
to the space ${\cal P}' \subset {\cal P}$ of 
local formal integrated power series independent of 
$(z,w)$, whenever $(z,w)$ form a set of coupled 
doublets under $\rho$, i.e.
\begin{eqnarray}
\rho z = w \, , ~~~~ \rho w =0 \,
\label{db}
\end{eqnarray}
and 
\begin{eqnarray}
[{\cal N},\rho] \neq 0 \, ,
\label{nc}
\end{eqnarray}
where ${\cal N} = z \frac{\delta}{\delta z} + w \frac{\delta}{\delta w}$
is the counting operator for $(z,w)$.
We notice that no restriction on the ghost number of the functionals
in ${\cal P}$ and ${\cal P}'$ is made. It might also happen that
${\cal P}$ and ${\cal P}'$ split into a sum of subspaces with different
ghost number.

This extends the well-known result \cite{Piguet:er,Barnich:2000zw}
stating that 
the cohomology of an arbitrary nilpotent differential $\rho$ is independent
of the set of variables $(z,w)$, provided that they form a set
of decoupled doublets, i.e. provided that, in addition to eq.(\ref{db}),
the following commutation relation holds:
\begin{eqnarray}
[{\cal N},\rho] = 0 \, .
\label{a6_1}
\end{eqnarray}

By making use of the results of
\cite{Barnich:db,Barnich:mt,doppietti}
 we can hence restrict ourselves to the analysis
of the cohomology of the restriction ${\cal S}'_0$ of
${\cal S}_0$ to the space of Lorentz-invariant local formal power
series independent of $(\lambda,\chi)$. We denote this space
by ${\cal F}'$.

${\cal S}'_0$ is explicitly given by
\begin{eqnarray}
\!\!\!\! {\cal S}_0' & = & {\rm Tr} \int d^3x \, \left ( (\partial_\mu \omega 
+ g[A_\mu,\omega] + c_\mu) \frac{\delta}{\delta A_\mu} 
+ \left . \frac{\delta \G^{(0)}}{\delta \bar c^{\mu *}} 
  \right |_{\lambda=\chi=0} \frac{\delta}{\delta \bar c^\mu} 
  \right . \nonumber \\
& & \left . + (\partial_\mu \omega + g[A_\mu,\omega] + c_\mu)
           \frac{\delta}{\delta \bar c^{\mu *}} 
	   - g \{c_\mu,\omega \} \frac{\delta}{\delta c_\mu}
	   - \frac{g}{2} \{ \omega, \omega \} \frac{\delta}{\delta \omega}
\right ) \, .
\label{wz1}
\end{eqnarray}
It is convenient to change generators according to
\begin{eqnarray}
&& c'_\mu = c_\mu + \partial_\mu \omega + g[A_\mu,\omega] \, , 
~~~~~ A'_\mu = A_\mu - \bar c^{*}_\mu \, .
\label{wz2}
\end{eqnarray}
${\cal S}_0'$ reads in the new variables:
\begin{eqnarray}
{\cal S}_0' & = & {\rm Tr} \int d^3x \, \left (
c'_\mu \frac{\delta}{\delta \bar c^{*}_\mu}  + \left . \frac{\delta \G^{(0)}}{\delta \bar c^{\mu *}} 
  \right |_{\lambda=\chi=0} \frac{\delta}{\delta \bar c^\mu}
- \frac{g}{2} \{ \omega, \omega \} \frac{\delta}{\delta \omega}
\right ) \, .
\label{wz3}
\end{eqnarray}
$(\bar c^{*}_\mu,c'_\mu)$ are again a set of coupled 
doublets under ${\cal S}_0'$
and by the same arguments used before the cohomology
$H({\cal S}'_0, {\cal F}')$ 
turns out to be isomorphic to  the cohomology of the
restriction ${\cal S}''_0$ of ${\cal S}'_0$ to the subspace ${\cal F}''$
of Lorentz-invariant integrated local formal power series
independent of $\bar c^{*}_\mu,c'_\mu$.
Notice that in this subspace $A'_\mu=A_\mu$. 
${\cal S}_0''$ has the following form: 
\begin{eqnarray}
{\cal S}_0'' = {\rm Tr} \int d^3x \Big ( - \left . 
\frac{\delta \Sigma'}{\delta \hat A^{\mu}} 
  \right |_{\scriptstyle{\pmatrix{\lambda=\chi=0, \cr \hat A_\mu = A_\mu , \cr
  c'_\mu=\bar c^{*}_\mu=0\cr}}} \frac{\delta}{\delta \bar c^\mu}
- \frac{g}{2} \{ \omega, \omega \} \frac{\delta}{\delta \omega}
\Big ) \, .
\label{wz4}
\end{eqnarray}
We have now to study the cohomology $H({\cal S}_0'',{\cal F}'')$
 of ${\cal S}_0''$ in the subspace
${\cal F}''$ spanned by Lorentz-invariant 
monomials generated by $A_\mu,\omega,\bar c_\mu$ and their derivatives.
For that purpose we use 
a very general result in cohomological algebra 
\cite{Piguet:er,Barnich:2000zw} stating that the
cohomology $H({\cal S}_0'',{\cal F}'')$ is isomorphic to a subset of 
the cohomology
$H({\cal L},{\cal F}'')$ of the Abelian approximation ${\cal L}$ to 
${\cal S}_0''$, which is given by
\begin{eqnarray}
{\cal L} = \int d^3x \, \left ( - \G^{ab}_{\mu\nu} A^b_\nu 
                             \frac{\delta}{\delta {\bar c}^{\mu a}} \right ) \, .
\label{wz5}
\end{eqnarray}
In the above equation $\G^{ab}_{\mu\nu}$ is the 1-PI classical 
two point-function,
defined by
\begin{eqnarray}
\G^{ab}_{\mu\nu}(x-y) = \left . \frac{\delta ^2 \Sigma'}{\delta A^a_\mu(x)
\delta A^b_\nu(y)} \right |_{\lambda=0, A_\mu=0} \, .
\label{wz6}
\end{eqnarray}

In order to study the cohomology of ${\cal L}$ we make use of the relationship
between ${\cal L}$ and the Koszul-Tate differential $\delta_{KT}$ \cite{Barnich:2000zw}
associated with the classical action
\begin{eqnarray}
S = - {\rm Tr} \int d^3x \, \left ( +\frac{1}{2} 
\epsilon^{\mu\nu\rho} A_\mu \partial_\nu A_\rho -
\frac{\alpha}{2} ( \partial A)^2 \right ) \, . 
\label{kt1}
\end{eqnarray}
The Koszul-Tate differential $\delta_{KT}$ acts in the space ${\cal F}_{KT}$
of local formal power series in $A_\mu,\omega$, 
their antifields $\bar c_\mu,\omega^*$ (with $\bar c_\mu$ playing the r\^ole of the antifield for $A_\mu$)
and their derivatives. The action of $\delta_{KT}$ on the generators of ${\cal F}_{KT}$ 
is the following:
\begin{eqnarray}
&& \delta_{KT} A_\mu^a = {\cal L} A_\mu^a = 0 \, , ~~~~ \delta_{KT} \omega^a = {\cal L} \omega^a = 0 \, , 
\nonumber \\
&& \delta_{KT} \bar c^a_\mu = {\cal L} \bar c^a_\mu = \frac{\delta S}{\delta A_\mu^a} \, , 
~~~~ \delta_{KT} \omega^{*a} = - \partial^\mu \bar c^{a}_\mu \, .
\label{kt_new1}
\end{eqnarray}
From the above equation we see that ${\cal L}$ coincides with the restriction of $\delta_{KT}$
to the subspace ${\cal F}'' \subset {\cal F}_{KT}$  of $\omega^*$-independent elements.

We define the antifield number as follows: $A_\mu^a$ and $\omega^a$
carry zero antifield number, $\bar c^a_\mu$ carries antifield number $+1$
and $\omega^{*a}$ carries antifield number $+2$.\footnote{Since with these
assignments $\delta_{KT}$ acts as a boundary operator (it has
degree $-1$ with respect to the antifield degree) the groups 
$H_m(\delta_{KT},{\cal F}_{KT})$ in antifield degree $m$
are said ``homology groups'' rather than ``cohomology groups'' of 
the differential $\delta_{KT}$.}

The general results on the homology groups of Koszul-Tate differentials
\cite{Barnich:2000zw} allow us to conclude that the homology groups
$H_m(\delta_{KT},{\cal F}_{KT})$ are zero in strictly positive
antifield number $m$. In particular
$H(\left . \delta_{KT} \right |_{{\cal F}''}, {\cal F}'') =
H({\cal L},{\cal F}'')$ must be $\bar c^\mu_a$-independent.
We remark that in general $H({\cal L}, {\cal F}'')$ is however
$A_\mu^a$-dependent \cite{Barnich:2000zw}.

By using the fact that $H({\cal L},{\cal F}'')$ must be 
$\bar c^\mu_a$-independent we can then further restrict
the analysis to the space ${\cal F}'''$ of Lorentz-invariant
integrated local formal power series spanned by $(\omega, A_\mu)$ and
their derivatives.
That is, the cohomology $H({\cal S}_0,{\cal F})$
is isomorphic to a subset of the cohomology 
$H({\cal S}_0''',{\cal F}''')$ of the restriction ${\cal S}'''_0$
of ${\cal S}''_0$ to the space ${\cal F}'''$:
\begin{eqnarray}
{\cal S}_0''' = {\rm Tr} \int d^3x \Big ( 
- \frac{g}{2} \{ \omega, \omega \} \frac{\delta}{\delta \omega}
\Big ) \, .
\label{wz9}
\end{eqnarray}

We now specialize to the sector ${\cal F}'''_{+1}$ 
of ${\cal F}'''$ with ghost number $+1$.
Any element $Y \in H({\cal S}'''_0,{\cal F}'''_{+1})$
must contain just one $\omega$ and must be 
${\cal S}'''_0$-invariant. It follows that it is zero, and then 
we can conclude that  $H({\cal S}'''_0,{\cal F}'''_{+1})$  is empty.
Going back to the cohomology of the full linearized classical ST
operator ${\cal S}_0$ we obtain that $H({\cal S}_0,{\cal F}_{+1})$
is also empty. Equivalently, eq.(\ref{a4}) holds. 
This ends the proof that the ST identities 
can be restored also at the $n$-th order
in the loop expansion.

\section{Consequences of the ST identities for the Green functions}\label{sec5}

In this section we analyze the consequences 
 of the ST identities
\begin{eqnarray}
{\cal S}(\G) =0 \, 
\label{c1}
\end{eqnarray}
for the Green functions of the model.
We remark that we can introduce in $\Sigma'$ (and correspondingly in the classical action in eq.(\ref{e11}))
a set of external sources 
$\zeta_i(x)$, coupled to BRST-invariant local composite operators
${\cal Q}_i(x)$, without spoiling the validity of the ST identities
in eq.(\ref{c1}).
The ST identities are also not broken  if we add a set of external sources 
$\tau_j(x)$ coupled to local composite operators 
${\cal O}_j(\hat A(A,\lambda))$, which depend only on $\hat A(A,\lambda)$.
Notice that the operators 
${\cal O}_j$ are BRST-invariant only modulo the equation
of motion for $\bar c_\mu$.

We use the collective
notation $\beta = \{ \zeta_i, \tau_j \}$ to denote all the sources $\zeta_i,
\tau_j$.
The connected generating functional $W$ is defined by the Legendre
transform of $\G$ with respect to the quantum fields of the model:
\begin{eqnarray}
W[J_\mu,J_{{\bar c}_\mu},J_{c_\mu};\omega, \lambda,\chi,\bar c^{* \mu},\beta] = 
\G + {\rm Tr} \int d^3x \, \left ( J_\mu A^\mu + J_{{\bar c}_\mu}{\bar c}^\mu
+ J_{c_\mu} c^\mu \right ) \, .
\label{c2}
\end{eqnarray}
The ST identities in eq.(\ref{c1}) read for $W$
\begin{eqnarray}
&& \!\!\!\!\!
{\cal S}(W)  =  - {\rm Tr} \int d^3x \, \left \{ \left ( (\partial_\mu \omega + g [ \frac{\delta W}{\delta J_\mu}, \omega]
+ \frac{\delta W}{\delta  J_{c_\mu}} \right ) J^\mu + \frac{\delta W}{\delta {\bar c}^{\mu *}} J_{{\bar c}_\mu}
\right .
\nonumber \\
&& ~~~~~~~~~~~~~~~~~~~~~~~~ \left . + \frac{g}{2} \{ \omega, \omega \} \frac{\delta W}{\delta \omega} - g \{ \frac{\delta W}{\delta J_{c_\mu}}, \omega \}
J_{c_\mu} \right \}
 + \chi \frac{\delta W}{\delta \lambda} =0 \, .
\label{c3}
\end{eqnarray}
The ST identities in eq.(\ref{c3}) allow to control the dependence of the connected Green functions on $\lambda$
and in particular to prove that the Green functions of BRST-invariant operators are independent of $\lambda$
\cite{Piguet:er,Piguet:1984js}.
Moreover, they also allow to prove that the Green functions of the operators ${\cal O}_j(\hat A(A,\lambda))$,
which are $\tilde s$-invariant
only modulo the classical equation of motion of $\bar c_\mu$, 
are also independent of $\lambda$.
The proof is standard \cite{Piguet:er,Piguet:1984js,ET}
and is reported here for completeness.
We first take the derivative of eq.(\ref{c3}) with respect to $\chi$ and
to $\beta_{j_1}(x_1), \dots, \beta_{j_n}(x_n)$ 
and finally go on-shell by setting $J_\mu=J_{{\bar c}_\mu}=J_{c_\mu}=\bar c^{* \mu}=\beta=\omega=\chi=0$.
This yields
\begin{eqnarray}
\left . \frac{\delta^{(n+1)} W}{\delta \lambda \delta \beta_{j_1}(x_1) \dots \delta \beta_{j_n}(x_n)} 
\right|_{on-shell} = 0 \, ,
\label{c4}
\end{eqnarray}
stating the independence of the Green functions 
\begin{eqnarray}
G^{(n)}_{j_1,\dots,j_n}(x_1,\dots,x_n) \equiv \left . \frac{\delta^{(n)} W}{\delta \beta_{j_1}(x_1) \dots \delta \beta_{j_n}(x_n)}
\right |_{on-shell}
\label{c5}
\end{eqnarray}
of $\lambda$.
Hence $G^{(n)}_{j_1,\dots,j_n}$
 can be equivalently computed at $\lambda \neq 0$ and
in the limit $\lambda=0$.
Since $\lambda=\frac{1}{m}$, eq.(\ref{c4}) entails that the Green functions 
in eq.(\ref{c5}) are independent of the topological mass parameter $m$.

\section{Physical states}\label{sec6}

Due to the validity of the ST identities
in eq.(\ref{c1}), the asymptotic BRST operator $Q$ \cite{Curci:1976yb,Kugo:zq,becchi}, 
acting on the asymptotic states
of the theory, is a conserved charge.
The action of $Q$ on the asymptotic fields of the model 
is given in the present case by the following commutation relations:
\begin{eqnarray}
&& [Q, A^{(as)}_\mu(x)] = c^{(as)}_\mu \, , ~~~~~~~~ \{ Q, c^{(as)}_\mu \} =0 \, ,\nonumber \\
&& \{ Q, {\bar c}^{(as)}_\mu \} = 0 \, .
 \label{fis1}
\end{eqnarray} 
From the above equation we see that, unlike $\tilde s$,
$Q$ is nilpotent.
This follows from the fact that $Q$ acts on on-shell fields.

The existence of a nilpotent conserved asymptotic BRST charge $Q$
allows us to apply 
the conventional BRST quantization of gauge theories
\cite{Weinberg:kr, becchi} to the present model.
The 
physical Hilbert subspace is defined as usual
 by ${\cal H}_{phys} = 
{\rm Ker \, } Q / {\rm Im \, } Q$. 
The vacuum is $Q$-invariant.
Moreover, we are only interested in
cohomology classes with zero ghost number.
From eq.(\ref{fis1}) we see that 
the pair $(A^{(as)}_\mu,c^{(as)}_\mu)$ forms
a $Q$-doublet and hence it does belong to ${\cal H}_{phys}$.
The only states belonging to ${\cal H}_{phys}$ are those
generated by ${\bar c}^\mu$, which, however, correspond to cohomology
classes with negative ghost number.

We conclude that no physical states with zero ghost number 
exist in this theory,
apart from the vacuum.
This is a well-known result for three-dimensional Chern-Simons theory: 
with the exception of the vacuum, there are no
physical states in the usual sense of perturbative field theory
\cite{Guadagnini:1989kr}. 
The non-trivial physical Hilbert space of Chern-Simons theory,
as described for instance in \cite{Witten:1988hf,Birmingham:1991ty} for the case of the three-manifold
$\Sigma \otimes R$ with $\Sigma$ a Riemann surface of non-trivial genus,
is beyond the purely perturbative framework of the present
analysis and thus cannot be dealt with by the cohomological
arguments of this section.

We point out that ordinary topological massive Yang-Mills theory,
without the inclusion of the antighost-ghost fields $(\bar c_\mu,c_\mu)$,
does possess local excitations.
The correspondence at the asymptotic level between the quantum theory
at $\lambda =0$ and at $\lambda \neq 0$ is actually preserved 
only thanks to the presence of the antighost-ghost fields $(\bar c_\mu,c_\mu)$.

We remark that our approach has the virtue to explicitly show
by purely cohomological arguments the 
absence in this model
of asymptotic states different from the vacuum, which
are physical in the sense of conventional perturbative
quantum field theory.
It is the structure of the BRST asymptotic charge $Q$ that enforces
the triviality of ${\cal H}_{phys}$ in the sector with zero ghost number.
In this construction the antighost-ghost fields, needed to take into
account the Jacobian of the gauge field redefinition in eq.(\ref{e4}),
play an essential r\^ole. This in turn  shows their relevance 
also at the asymptotic level for a proper 
quantum implementation of the classical equivalence between
three-dimensional Chern-Simons theory and
topological massive Yang-Mills theory.

\section{Conclusions}\label{sec7}

In this paper we have analyzed the extension to the quantum level of  
the classical equivalence upon a redefinition of the gauge connection
between Chern-Simons theory and topological massive Yang-Mills theory
in three-dimensional Euclidean space-time,
within the BRST formulation of the Equivalence Theorem. 
The parameter controlling the change in the gauge connection 
is the inverse $\lambda$ of the topological mass $m$.  
At the quantum level one needs to take into account the Jacobian
of the gauge field redefinition in eq.(\ref{e4}) in order to
properly extend the classical equivalence to the full
$\lambda$-dependent quantum effective action.
This can be done in a local and covariant way by introducing 
a set of antighost and ghost fields $\bar c^\mu, c^\mu$ with the
same Lorentz-vector index as $A_\mu$.
The BRST differential $\tilde s$ associated with the gauge field
redefinition has then been derived and the corresponding Slavnov-Taylor (ST) 
identities have been proven to be anomaly-free.
$\tilde s$ incorporates both the BRST differential $s$, issued from
the gauge invariance of the ungauged topological Chern-Simons theory,
and the BRST differential $\delta$, associated with the field
redefinition in eq.(\ref{e4}).
As a consequence of the ST identities, 
the Green functions of local operators depending only 
on the transformed
gauge connection $\hat A(A,\lambda)$, 
as well as those of BRST invariant operators,
are shown to be $\lambda$-independent.
Hence they can be equivalently computed at $\lambda \neq 0$ and
in the limit $\lambda =0$. Since $\lambda=\frac{1}{m}$, this entails
that such a  selected set of Green functions is independent of 
the topological mass $m$.
The identification of the physical states of the model within
the conventional perturbative approach to gauge theories can be 
carried out by following the standard BRST cohomological construction.
It turns out that the antighost and ghost fields $\bar c^\mu,c^\mu$
play an essential r\^ole in this
procedure, trivializing the cohomology of the asymptotic BRST charge 
in the sector with zero ghost number.
This guarantees that the correspondence at the level of asymptotic
states between the quantum theory at $\lambda =0$ and at $\lambda \neq 0$  
is actually preserved.
This in turn provides an additional insight into the relevance of 
the antighost and ghost fields, associated with the gauge field redefinition,
for a proper quantum implementation of 
the classical equivalence between Chern-Simons
theory and massive topological Yang-Mills theory
in three-dimensional flat Euclidean space-time.

\section*{Acknowledgments}

Useful discussions with Prof.~S.~P.~Sorella are gratefully acknowledged.

\end{document}